\documentclass[sigconf]{acmart}

\AtBeginDocument{%
  \providecommand\BibTeX{{%
    \normalfont B\kern-0.5em{\scshape i\kern-0.25em b}\kern-0.8em\TeX}}}

\setcopyright{none}
\renewcommand\footnotetextcopyrightpermission[1]{} 
\usepackage{xcolor}
\usepackage{algorithmic}
\usepackage{graphicx}
\usepackage{textcomp}
\usepackage{xcolor}
\usepackage{listings}
\usepackage{multirow}
\usepackage{mathdots}
\usepackage{tikz}
\usepackage{siunitx}
\usepackage{makecell}
\usetikzlibrary{shapes}
\usetikzlibrary{positioning}
\usepackage[inline]{enumitem}
\usepackage{balance}
\usepackage{algorithm2e}
\usepackage{listings}
\usepackage{circledsteps}

\newcommand*\annotatedFigureBoxCustom[8]{\draw[#5,thick,rounded corners] (#1) rectangle (#2);\node at (#4) [fill=#6,thick,shape=circle,draw=#7,inner sep=2pt,font=\sffamily,text=#8] {\textbf{#3}};}
\newcommand*\annotatedFigureBox[4]{\annotatedFigureBoxCustom{#1}{#2}{#3}{#4}{black}{white}{black}{black}}

\newenvironment {annotatedFigure}[1]{\centering\begin{tikzpicture}
\node[anchor=south west,inner sep=0] (image) at (0,0) { #1};\begin{scope}[x={(image.south east)},y={(image.north west)}]}{\end{scope}\end{tikzpicture}}

\definecolor{WhiteSmoke}{rgb}{0.96, 0.96, 0.96}
\definecolor{DarkRed}{rgb}{0.55, 0.0, 0.0}
\definecolor{darkolivegreen}{rgb}{0.33, 0.42, 0.18}
\definecolor{mediumspringbud}{rgb}{0.89, 0.96, 0.64}
\definecolor{officegreen}{rgb}{0.0, 0.5, 0.0}
\definecolor{orange-red}{rgb}{1.0, 0.27, 0.0}

\lstdefinestyle{verilog} {
    language=Verilog,
    morekeywords={unique, logic, always_comb, typedef, enum},
}

\definecolor{diffstart}{named}{WhiteSmoke}
\definecolor{diffincl}{named}{officegreen}
\definecolor{diffrem}{named}{orange-red}

\lstdefinelanguage{diff}{
  basicstyle=\ttfamily\small,
  morecomment=[f][\color{diffstart}]{@@},
  morecomment=[f][\color{diffincl}]{+\ },
  morecomment=[f][\color{diffrem}]{-\ },
}

\newcommand{\hgdb}{\texttt{hgdb}{ }}

\begin{document}

\title{Bringing Source-Level Debugging Frameworks to
Hardware Generators
}

\author{Keyi Zhang}
\email{keyi@cs.stanford.edu}
\affiliation{
  \institution{Stanford University}
  \city{Stanford}
  \state{California}
  \country{USA}
}
\author{Zain Asgar}
\email{zasgar@stanford.edu}
\affiliation{
  \institution{Stanford University}
  \city{Stanford}
  \state{California}
  \country{USA}
}
\author{Mark Horowitz}
\email{horowitz@ee.stanford.edu}
\affiliation{
  \institution{Stanford University}
  \city{Stanford}
  \state{California}
  \country{USA}
}

\acmConference[DAC’22]{Design Automation Conference}{July 10-14}{San Francisco, CA, USA}

\begin{abstract}
  High-level hardware generators have significantly increased the productivity of design engineers.
They use software engineering constructs to reduce the
repetition required to express complex designs and enable more composability. However, these benefits are
undermined by a lack of debugging infrastructure, requiring hardware designers to debug generated, usually incomprehensible, RTL code.
This paper describes a framework that connects modern software source-level debugging frameworks to RTL created from hardware generators. 
Our working prototype offers an Integrated Development Environment (IDE) experience for generators such as RocketChip (Chisel), allowing
designers to set breakpoints in complex source code, relate RTL simulation state back to source-level variables, and do forward and backward debugging, with almost no simulation overhead (less than 5\%).

\end{abstract}



\maketitle

\section{Introduction}
The past two decades have seen many hardware generation frameworks (HGFs) that increase design
productivity by mixing software programming languages with hardware description languages (HDLs).
Either a scripting language is added on-top of an existing HDL~\cite{genesis} or the HDL is embedded into another
high-level programming language, such as Chisel (Scala)~\cite{chisel}, Mamba~\cite{mamba}, and Magma (Python)~\cite{magma}.
A growing trend in this hardware generator domain is to apply software engineering concepts 
such as object-oriented programming and functional programming to hardware design, which has been shown to increase design
productivity~\cite{chisel}.

However, even though generators have moved the hardware description up a level closer to software engineering, many of the IP blocks used in
the system are still coded/transferred at the RTL level, and all state of the art logic simulators still focus on this level.
This reality forces HGF designers to debug system and interface errors through RTL simulation.
Our solution to this problem lets the programmer leverage advanced
software debugging techniques, developed over decades, to debug
hardware designs simulated at the RTL level using the context of a generator's source code and data structures.
Just like everyone expects to debug their C-code using a source-level debugger, 
we think hardware designers deserve the same type of debugging environment, without having to reason about the generated RTL, i.e. the assembly language.


The debugging system should be orthogonal to the testing environment so that the framework can be used with well-established
testing frameworks such as Universal Verification Methodology (UVM). This allows designers to leverage drivers and monitors provided
by testing frameworks that have been widely used in practice and shown to reduce the verification difficulties
on large systems~\cite{uvm}. Thus our new system can combine the benefits of software debugging and hardware testing frameworks.
In addition, because simulators are capable of capturing the traces for every state change, we can replay the trace and
enable reverse-debugging, which is much more challenging to implement for software.

This paper presents our system, \texttt{hgdb} (\textbf{h}ardware \textbf{g}enerator \textbf{d}e-\textbf{b}ugger),
which is capable of debugging designs
from various HGFs using commercial simulators and trace files. Key contributions of this paper include:
\begin{enumerate*}
    \item A modular software architecture with a well-defined interface for source-level debugging that allows existing tools to cleanly and efficiently interact.
    \item A breakpoint emulation approach with minimal simulation overhead.
    \item A prototype implementation of this type of debugger, \texttt{hgdb}, that allows source-level debugging for Chisel on
    complex designs.
\end{enumerate*}

\section{Background}
In the early stages of software engineering, developers had to write their code in a high-level language such as C or COBOL and
then debug the assembly code directly after compiler optimizations, which is difficult to do.
David Ditzel and David Patterson~\cite{high-level-language} argued that
the computer system should use high-level languages for both programming and debugging, and the system should
report execution errors in terms of the high-level language source program. To enable high-level debugging,
the compiler needs to track the source file information and store it in a separate \textit{symbol table}
that maps variable names to actual memory addresses and vice versa.
During program execution, an external tool, a \textit{debugger}, can load this symbol table and display the values with their
original variable names instead of memory addresses. For a program where variables have different lifetimes, the
variable binding, or scope, depends on the context of execution, and the debugging mapping must change
accordingly. One major task for debuggers is to construct the scoped stack frame information based on the symbol table when
required.

To allow easy interoperability between tools, software debuggers have adopted a set of standards, such as DWARF~\cite{dwarf}.
Regardless of the targeted programming languages, these systems essentially provide a mechanism that maps source-level
information to a construct understandable by the debugger. The mapping includes
\begin{enumerate*}[label=(\alph*)]
  \item variable and scope mapping; and
  \item line number mapping.
\end{enumerate*}

On the other hand, although HGFs such as Chisel and PyMTL have their own performant simulators that directly digest their respective intermediate representations (IRs)~\cite{ESSENT, mamba}, they do not offer any source-level debugging support, nor do they incorporate external IPs written in conventional Verilog. As a result, designers still have to resort to traditional waveforms with somewhat obfuscated RTL when debugging system-level issues.

Inspect~\cite{calagar2014source} is an experimental system designed for debugging High-Level Synthesis (HLS) on FPGA.
Inspect defines a symbol table which is used to reconstruct the source execution state so users can debug HLS programs at source-level. Because it is tightly integrated into the compiler tool chain, it has control over the code generation as well as various compiler passes. To facilitate debugging, Inspect inserts additional circuits into the underlying design and gathers debugging information during FPGA emulation. As a result, Inspect is tightly locked to its own compiler tool chain and a specific FPGA vendor.
Our proposed system is inspired by Inspect, but instead focuses on traditional RTL simulations. It also decouples the dependencies on compilers and testing environment and avoids inserting additional RTL logic into the design to minimize undesirable simulation overhead.

Industrial debugging tools such as Verdi~\cite{verdi} employ powerful debugging techniques that enable hardware designers to quickly identify bugs in the RTL code. However, they are less useful for HGFs since the generated RTL does not convey the design intent, as shown in Listing~\ref{lst:bug-code-chisel} and ~\ref{lst:bug-code-rtl}. Our work addresses the challenge to correlate the problematic signals back to the source code and provide source-level debugging.

\section{\hgdb: A Debugger for HGF's}\label{sec:hardware-debug}
Building an \textit{ambitious} system that supports various simulation environments with a unified debugging infrastructure is clearly challenging. Many software debugging frameworks instead opt for language and platform specific debugging environments. To achieve our design goals, we believe the key is to define a clean interface between our three main components: simulators, debuggers, and symbol table. 

One challenge is how to implement breakpoints: only a few simulators
allow users to set breakpoints interactively, and their interface varies. To support all simulators means \texttt{hgdb} will need to emulate the breakpoints externally to the simulators; emulating breakpoints faithfully and efficiently is critical.

Another potential challenge is that developers rarely use an HGF to generate the entire design and test bench; instead, IP blocks from different sources are composed together to form a larger system and then simulated under a complex testing environment. As a result, \texttt{hgdb} only has a partial view of the final design and it needs a method to locate the generated IP in the complete system during simulation.

To address the breakpoint challenge, we leverage two facts about modern designs
\begin{enumerate*}
    \item most designs are synchronous, which implies that signals have to be stable before the rising edge of the clock;
    \item most RTL simulations use zero-delay logic models.
\end{enumerate*}
The first means that our system only needs to check for potential breakpoints \textit{at} the rising edge of the clock. The second means that all logical values will be stable at every clock edge in the simulator. Thus, for a small loss in efficiency, one does not need to associate breakpoints with clock domains: one can check all breakpoints on every clock transition.  Of course, evaluating breakpoints only on clock transitions create a number of subtle issues which \texttt{hgdb} must address.  These issues are discussed in Section~\ref{sec:breakpoint}.


In terms of efficiency, for a large scale simulation,
most of the computation time within a simulated clock cycle is consumed on evaluating logic performed in this cycle. Compared to this time intensive task, calling into a custom routine at each simulated clock cycle results in low overhead (5\% as shown in Section~\ref{sec:benchmark}).

To locate the generated IP within the entire design, we use the fact that all simulators have interfaces to query design hierarchy and hierarchical signal names. 
Since most HGFs use a number of compiler passes to lower some form of IR to RTL, these passes can be used to extract out symbol mappings.
Although the design hierarchy presented in this symbol table is only a subset of the final design, the relative hierarchy does not change. Section~\ref{sec:mapping} explains how \texttt{hgdb} can locate the generated IP block and use it to map source signals to the actual RTL hierarchical path inside the test bench by finding the block with matching module/signal names.

\begin{figure}[htb]
  \centering
  \includegraphics[width=0.95\linewidth]{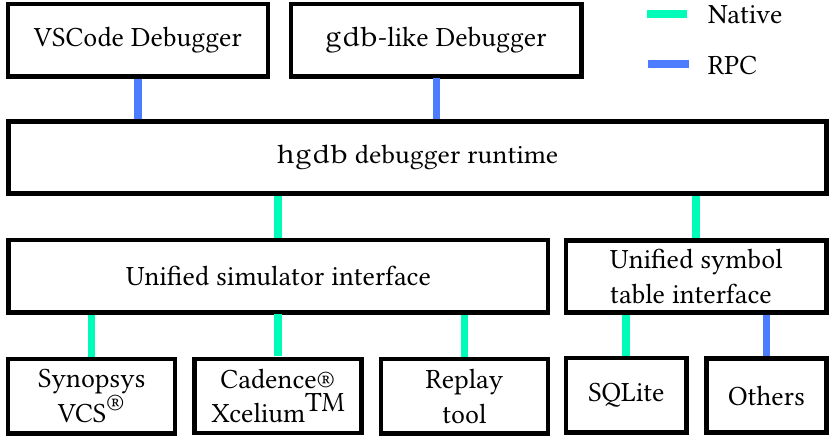}
  \caption{The \hgdb runtime is loaded into simulator tools natively via a unified simulator interface.
  Debugger tools communicate with the runtime using RPC protocol, and the symbol table can be either queried natively or via RPC.}\label{fig:overview}
  \Description[\hgdb system overview]{\hgdb system overview}
\end{figure}

As a result, \texttt{hgdb} defines two sets of unified interfaces: one for the simulator, and another one for the symbol table, as shown in Figure~\ref{fig:overview}. Each interface defines a small set of primitives required to implement the functionality described above. By abstracting away the vendor-specific APIs and implementation details, we can maximize the compatibility across different tools and systems,
which we believe is unique among hardware debugging systems. In addition, since the debugger runtime directly interacts with the simulator, we use native implementation to ensure minimal simulator
overhead. However, for debugging interactions that are less timing sensitive, we use Remote Procedure Call (RPC) to enable compatibility.

\subsection{Emulating Breakpoints in Simulation}\label{sec:breakpoint}
The challenge in using state values after the simulator reaches equilibrium is that some desirable intermediate values are overwritten while the
state is converging. This typically happens when the same variable is reused inside combinational
logic to obtain the final result. For instance, a variable called \texttt{sum} can be overwritten multiple
times inside a \texttt{for} loop for accumulation. If we inspect the design state at the next clock edge, we will only see the final result and lose all the intermediate partial sums, as shown by a C-like pseudo code in Listing~\ref{lst:ssa-before}.

Our solution is to leverage a widely used algorithm in HGF compilers, namely Static Single Assignment (SSA)~\cite{ssa}.
Because the symbol name aliasing problem only happens
in combinational logic,
SSA is the perfect solution.
During the SSA transform, fixed-length loops get unrolled and conditional statements get flattened so that
each variable will only be assigned exactly once. Listing~\ref{lst:ssa-after} shows the pseudo-code after the SSA transformation.
Notice that the transform creates several temporal variables to hold the value of \texttt{sum},
which implies that the variable mapping can be different depending on context. We can use standard compiler techniques to track
the variable mapping. In this case, if the breakpoint hits Line~\ref{line:ssa-after-1} in Listing~\ref{lst:ssa-after},
we should fetch the value of \texttt{sum0} to represent \texttt{sum}, and \texttt{sum1} at Line~\ref{line:ssa-after-2}.

Due to loop unrolling and SSA, if the user sets a breakpoint
at Line~\ref{line:ssa-before-1} in Listing~\ref{lst:ssa-before}, we need to emulate two breakpoints in Listing~\ref{lst:ssa-after}, one at Line~\ref{line:ssa-after-1}
and one at Line~\ref{line:ssa-after-2}. However, since these two mapped statements always execute regardless of the input condition,
we need something called an \textit{enable} condition that determines which line can be active during simulation. For instance, the enable condition for Line~\ref{line:ssa-after-1} in Listing~\ref{lst:ssa-after} is \texttt{data[0] \% 2}, which specifies that
the potential breakpoint can be enabled only if \texttt{data[0]} is odd.
The \texttt{enable} condition can be obtained by AND-reduction
on the SSA transform condition stack. For conditional logic in sequential blocks, \texttt{enable}
condition extraction is much easier, since we can simply walk up the hierarchy in the syntax tree and perform an AND-reduction on the
conditions.

\begin{figure}[!htb]
\begin{lstlisting}[language=C++,
  caption={\texttt{for}-loop using \texttt{sum} to accumulate \texttt{data}.},
  label=lst:ssa-before,
  captionpos=b,
  numbers=left,
  stepnumber=1,
  escapechar=!]
int sum = 0;
for (int i = 0; i < 2; i++) {
  if (data[i] % 2)
    sum += data[i];!\label{line:ssa-before-1}!!\tikz[remember picture] \node [] (a){};!
}
\end{lstlisting}
\begin{tikzpicture}[remember picture, overlay,
  every edge/.append style = { ->, thick, >=stealth,
                                dashed, line width = 1pt },
  every node/.append style = { align = center, minimum height = 10pt,
                               fill= mediumspringbud} ]

\node [right = .6cm of a]  (A) {\scriptsize Multiple line-mapping after SSA transform};
\draw (A.west) edge (a.east) ;
\end{tikzpicture}
\Description[C++ code before SSA transform]{C++ code before SSA transform}
\end{figure}

\begin{figure}[!htb]
\begin{lstlisting}[language=C++,
  caption={Summation accumulation transformed after loop unrolling and SSA.},
  label={lst:ssa-after},
  captionpos=b,
  numbers=left,
  stepnumber=1,
  escapechar=!]
int sum, sum0, sum1, data[2], data0, data1;
sum0  = 0;
data0 = data[0] % 2? data[0]: 0;
sum1  = sum0 + data0; !\tikz[remember picture] \node [] (a){};! !\label{line:ssa-after-1}!
data1 = data[1] % 2? data[1]: 0;
sum2  = sum1 + data1; !\tikz[remember picture] \node [] (b){};! !\label{line:ssa-after-2}!
sum   = sum2;
\end{lstlisting}
\begin{tikzpicture}[remember picture, overlay,
  every edge/.append style = { ->, thick, >=stealth,
                                dashed, line width = 1pt },
  every node/.append style = { align = center, minimum height = 10pt,
                               fill= mediumspringbud} ]

\node [right = .6cm of a]  (A) {\scriptsize Enable condition: \texttt{data[0] \% 2}};
\node [right = .6cm of b]  (B) {\scriptsize Enable condition: \texttt{data[1] \% 2}};
\draw (A.west) edge (a.east) ;
\draw (B.west) edge (b.east) ;
\end{tikzpicture}
\Description[SSA transformed code]{SSA transformed code}
\end{figure}

\subsection{Breakpoint Scheduling and Reverse-Debugging}
Since the breakpoints are emulated in software, \hgdb is free to choose the ordering in which breakpoints are evaluated.
If supported by the underlying simulator/toolchain, we can also use the same scheduling algorithm for reverse-debugging.

Notice that because hardware is concurrent, there could be tens of threads that share the same source information but operating on
different data. To faithfully represent this concurrent behavior, we insert and schedule all breakpoints that share the same source location requested
by the user.
This is similar to \textit{thread information} in software debugging, where users can select different threads during a breakpoint, as shown in
Figure~\ref{fig:ide}~\hyperlink{mark:b}{\Circled{\textbf{B}}}.

\begin{figure}[!htb]
  \centering
  \includegraphics[width=1\linewidth]{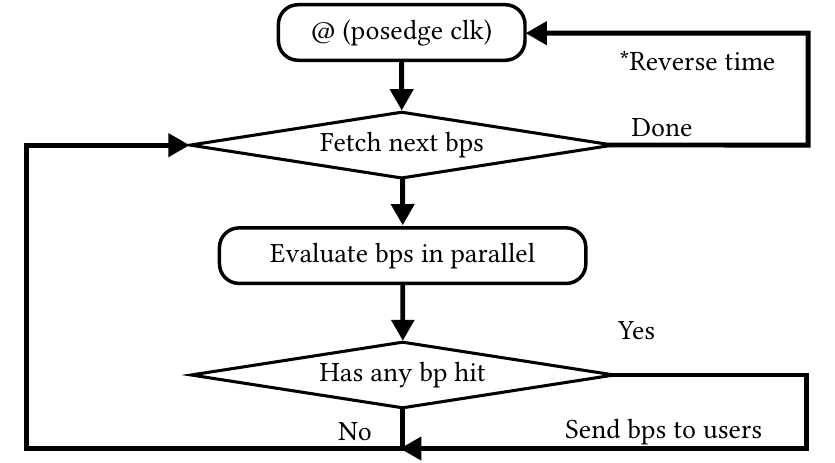}
  \caption{\hgdb breakpoints (bps) scheduling algorithm. The scheduling loop finishes when there is no breakpoint left to schedule.
  If the scheduler is in reverse-mode, we can reverse the time if the underlying simulator supports it.}\label{fig:bp-scheduling}
  \Description[\hgdb breakpoint scheduling algorithm]{\hgdb breakpoint scheduling algorithm}
\end{figure}

The scheduling algorithm is shown in Figure~\ref{fig:bp-scheduling}. Before the simulation starts, we compute the absolute ordering of every
potential breakpoint based on the symbol table. In most cases, they are ordered by their lexical order and scopes, i.e.
ordered by line and column number within a function. At the positive edge of the clock, we enter this breakpoint evaluation loop:
\begin{enumerate*}
  \item\label{enum:bp-schedule-step-1} we select a list of inserted breakpoints that share the same source location based on the pre-computed ordering. If there is no
  breakpoint left to select, we exit the loop and wait for next clock edge;
  \item we evaluate each breakpoint condition in parallel. Besides the breakpoint enable condition computed from SSA, each breakpoint
  can also contain conditional expressions specified by the user;
  \item if any breakpoint matches the condition, we reconstruct the stack frame based on the symbol table and then send the result to the user;
  \item once the user responds with a command, we loop back to step~\ref{enum:bp-schedule-step-1}.
\end{enumerate*}

There are several benefits of using such loop-based breakpoint scheduling given our breakpoint emulation mechanism. First, we can exit the loop
immediately if there is no breakpoint inserted, thus minimizing runtime overhead. Second, we can evaluate multiple breakpoints at the same
time, reducing latency between breakpoint hits in the runtime and when user sees the reconstructed frames. Third, which is more subtle, we
allow reverse-debugging within the same timestamp, which we call \textit{intra-cycle} reverse debugging, regardless of the underlying simulators. Notice that in a normal debugging
scenario, we select breakpoints in the order specified by the symbol table. If we reverse the selection order, however, we can create the
\textit{illusion} of going back in time as statements in the source code appear to be executed in reversed order. If the underlying simulator
supports reversing time, such as a trace-based replay engine, we can extend this reverse-debugging capability to its full potential, since we can go to previous clock cycle and start breakpoint selection in reversed order again.

\subsection{Unified Simulator Interface}
When selecting simulator interface primitives, one obvious challenge is how to support various simulator vendors, both commercial and open-source, and also
enable offline replay from captured trace. Based on our prototyping experiences, we defined a minimum set of simulator interface primitives that enable \texttt{hgdb} to function. 

\begin{itemize}
  \item \textbf{Get signal value}. This is essential to emulate breakpoint and reconstruct frames.
  \item \textbf{Get design hierarchy and clock information}. This is used to map generated RTL to the full test environment when
  the testbench is produced outside the HGFs. We also need the clock information to know when to pause the simulation.
  \item \textbf{Place callbacks on clock changes}. This is required to allow \hgdb to evaluate breakpoints.
  \item \textbf{Get and set simulation time}. This optional command enables reverse debugging.
  \item \textbf{Set signal value}. This optional command enables the debugger to change simulation values (not possible when interfacing with a trace file). 
\end{itemize}

These primitives are implemented using the Verilog Procedural Interface (VPI), an intrinsic part of the SystemVerilog
standard. To maximize the compatibility, \hgdb only uses a small subset of VPI functions, ones that are supported
by all simulator vendors, to implement these primitives. For tools that do not have a VPI interface,
we implement these primitives using additional design information. For instance, since VCD-based traces only contain design hierarchy but not definition information,
we can use instance names from the symbol to figure out the actual hierarchy mapping, using common substring matching.

The interface is handled automatically by an Application Binary Interface
(ABI) during executable linking time, which makes \hgdb easy to integrate into any existing simulator environment.

\subsection{Symbol Table}\label{sec:mapping}
Symbol tables are crucial for translating filename/line numbers into breakpoint and reconstructing stack frame information. 
Based on our experience with different HGFs, we select a minimum set of primitives that can be easily provided by each HGF:
   
\begin{itemize}
  \item \textbf{Get breakpoints from source location}. This is used to translate line information to the actual breakpoints.
  \item \textbf{Get scope information for each breakpoint}. This is required to construct a frame when a breakpoint is hit.
  \item \textbf{Resolve scoped variable names to RTL name}. When constructing frames for the breakpoint, \hgdb needs to translate source-level local variables
  into full hierarchical RTL names to query the simulator interface.
  \item \textbf{Resolve instance variable names to RTL name}. Similar to scoped variable name, \hgdb needs to translate instance-specific variables into full hierarchical RTL names.
\end{itemize}

\begin{figure}
\includegraphics[width=1\linewidth]{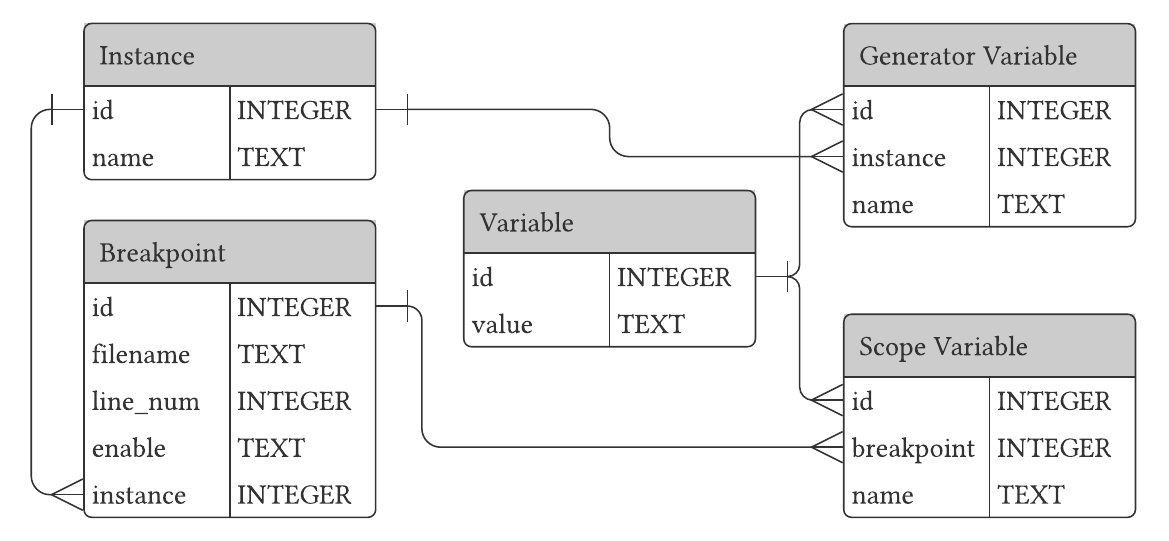}
\caption{Simplified schema diagram for SQLite-based symbol table. Arrows in the figure illustrate relations, which can be used to improve
search performance and guarantee data integrity.}
\label{fig:sql-schema}
\Description[hgdb schema diagram]{hgdn schema diagram}
\end{figure}

Symbol table primitives are queried either through RPC or ABI implemented via a native SQLite database.
The SQL schema is designed to be simple yet efficient to query debugging information, as shown in Figure~\ref{fig:sql-schema}.
Table \texttt{Instance} describes  hierarchical name in RTL and table \texttt{Breakpoint} encodes the source location
information as well as the enable condition. It also references the instance so we can reconstruct the generator instance variables once
a breakpoint hits. \texttt{Scope Variable} together with \texttt{Variable} can be used to construct the frame associated with any breakpoint.
The symbol table primitives can be naturally translated into relational queries since these entities are linked together.

Notice that since the simulator is paused whenever \hgdb interacts with the symbol table, e.g. either inserting
breakpoints or constructing frames, the symbol table performance is less important compared to the simulator interface.
This makes RPC-based implementations desirable for frameworks that implement their own symbol table.

\subsection{Debugger Environment}
\begin{figure}

  \begin{annotatedFigure}
    {\includegraphics[width=\linewidth]{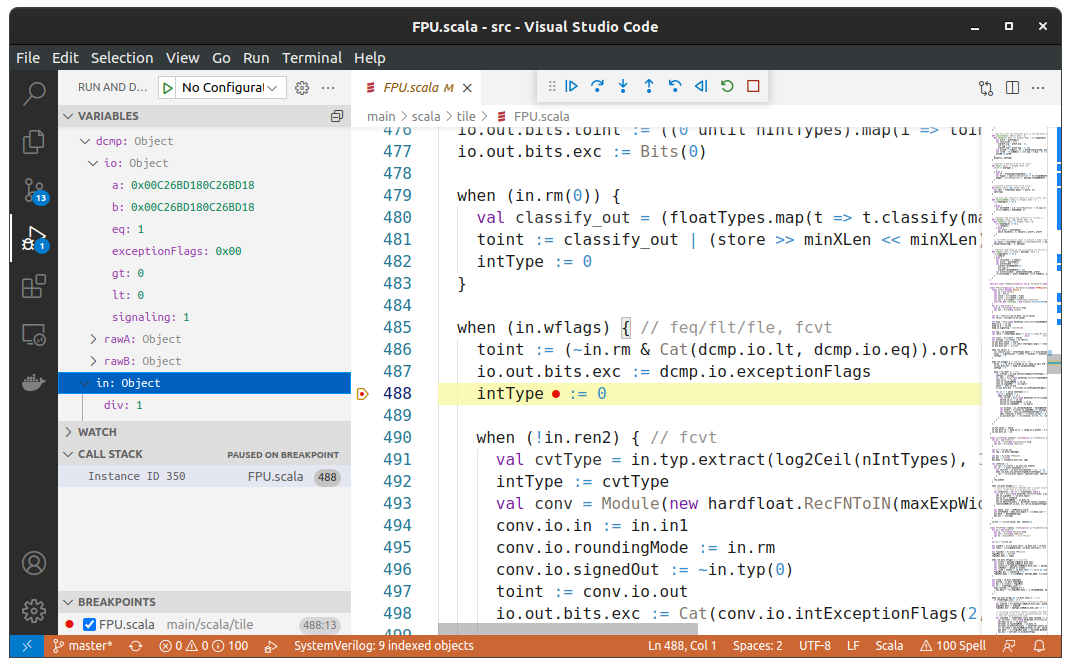}}
    \annotatedFigureBox{0.058,0.3738}{0.323,0.8162}{\hypertarget{mark:a}{A}}{0.323,0.8162}
    \annotatedFigureBox{0.057,0.2148}{0.33,0.3321}{\hypertarget{mark:b}{B}}{0.33,0.2148}
    \annotatedFigureBox{0.495,0.8362}{0.732,0.898}{\hypertarget{mark:c}{C}}{0.495,0.8362}
    \annotatedFigureBox{0.333,0.355}{0.865,0.475}{\hypertarget{mark:d}{D}}{0.865,0.355}
  \end{annotatedFigure}

  \caption{VSCode-based \hgdb IDE. \protect\hyperlink{mark:a}{\Circled{\textbf{A}}}~Values of local and generator variables fetched from the simulator.
  \protect\hyperlink{mark:b}{\Circled{\textbf{B}}} Concurrent hardware threads executing the same line.
  \protect\hyperlink{mark:c}{\Circled{\textbf{C}}}~Simulation state control such as continue, step over, and reverse-step.
  \protect\hyperlink{mark:d}{\Circled{\textbf{D}}}~Source- and conditional-breakpoints.}
  \label{fig:ide}

  \Description[VSCode screenshot]{VSCode screenshot}
\end{figure}
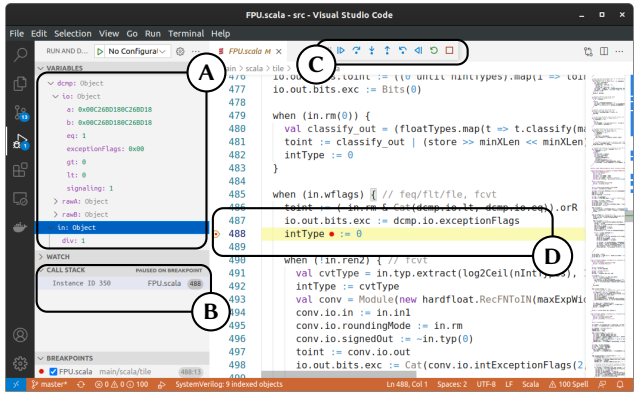

The debugger is one of the most important parts of the debugging experience. \hgdb offers two debuggers. One is a GNU Debugger (\texttt{gdb})-inspired
debugger, and the other is more akin to an IDE, implemented as a Visual Studio Code (VSCode) extension.
Users can directly view source code and set breakpoints in the IDE, and issue debugging commands. In order to support
different debuggers, \hgdb relies on RPC-based debugging protocol similar to \texttt{gdb remote protocol}, where the debugger
connects to \hgdb via WebSocket.
Figure~\ref{fig:ide} shows the \hgdb IDE, where users can write and debug code in the same environment.

\section{Case Study Using Chisel}\label{sec:case-study}
Chisel is one of the most popular HGFs, and it has consequently been widely used for hardware design~\cite{chisel}. Its frontend
language is embedded in Scala, which gets compiled into an IR called FIRRTL~\cite{firrtl}.
Then the design is further optimized and lowered to RTL. In this section, we first illustrate how to extract symbol table information
from the IR, and then demonstrate how we can use \hgdb to debug RocketChip, a popular RISC-V SoC written in Chisel, and
finally we benchmark the performance impact on simulation.

\subsection{Extracting Symbol Table from FIRRTL}
The FIRRTL IR has three levels, \texttt{High}, \texttt{Middle}, and \texttt{Low}. The \texttt{High} form has high-level constructs, such as port bundle, that must be lowered for synthesis. Designs written in Scala are typically
transformed into the \texttt{High} form of FIRRTL, which goes through multiple optimization passes and eventually gets lowered
to a much restricted set of primitives such as registers and multiplexers.

Due to the semantic gap between Scala and FIRRTL, Chisel
has to perform transforms to lower some high-level constructs into FIRRTL. For instance, a one-liner of \texttt{map} and
\texttt{reduce} chained calls in Chisel may result in tens of lines of FIRRTL statements, since FIRRTL, being closer to hardware
representation, is not functional. As a result, there are some transformation artifacts in the IR that make it difficult to
read. Furthermore, there are many optimization passes employed by default in FIRRTL, such as constant
propagation, common sub-expression elimination, and dead code elimination~\cite{firrtl}. Although these optimizations reduce
the final netlist gate count, they make the final RTL challenging to debug.

Nevertheless, there is still useful information we can extract from the \texttt{High} form of FIRRTL. For instance,
Chisel stores original Scala filenames and line numbers in FIRRTL for variable declaration and assignment, which can be used to compute breakpoints.
In addition, the \texttt{High} form of FIRRTL preserves high-level logic such as conditional statements, which allows us to compute the breakpoint's \texttt{enable} condition.

Since FIRRTL is an IR, we can write passes to extract the debugging information and compute symbol table for \texttt{hgdb}. FIRRTL does not offer
symbol tracking natively, so, in order to work
with compiler optimization, we adopt a two-pass approach. The first pass operates on the \texttt{High} form where the IR best resembles the source
structure, and the second pass operates on the \texttt{Low} form where the IR is close to generated RTL. As shown in Algorithm~\ref{alg:extract-firrtl},
the first pass annotates variables and statements of interest; the second pass collects these annotations and computes the final mapping and symbol
table. 
Function \texttt{ComputeSymbolTable} employs some heuristics to recover high-level information as much as possible given the FIRRTL information.
Since the annotation collection operates on the \texttt{Low} form, if the compiler optimization removes a variable, we will not see it in the \texttt{Low} form.
As a result, the generated symbol table will not contain the variable optimized away, a behavior consistent with software compilers. 
We also need to compute our enable condition in the \texttt{High} form because, once the IR
is lowered to \texttt{Low} form, FIRRTL flattens the structure and uses a \texttt{MUX} instead, making the condition extraction more difficult.

In debug mode, similar
to \texttt{gcc}'s \texttt{-O0}, the first pass can insert \texttt{DontTouchAnnotation}, which keeps the target IR node away from any compiler
optimization. This will bloat the generated RTL and slow down the simulation, since FIRRTL will not perform any optimization; however, it contains all the source information, which can be useful
for debugging. We have noticed about 30\% increase in the symbol table size when the debug mode is on.

\begin{algorithm}[tb]
  \caption{Extract symbol table from FIRRTL. In the first pass, we annotate every FIRRTL IR node; in the second pass, we compute
  the symbol table based on the current (potentially optimized) circuit state.}\label{alg:extract-firrtl}
  \KwIn{$CircuitState$}
  \KwOut{${Table}$}
  \tcp*[h]{First pass to annotate FIRRTL nodes}\;
  $Annotations \gets \{\}$\;
  \ForEach{$node \in CircuitState$} {
    \lIf{$node$ is statement} {
        $node.enable \gets ComputeEnableCondition(node)$\
    }
    $Annotation \gets Annotations \cup \{node\}$
  }

  \tcp*[h]{Second pass after FIRRTL transformations}\;
  $IRNodes \gets \{\}$\;
  \ForEach{$node \in Annotations$} {
    \lIf{$node \in CircuitState$} {
      $IRNodes \gets IRNodes \cup node$\
    }
  }

  $Table \gets ComputeSymbolTable(IRNodes)$\;
\end{algorithm}

\subsection{Debugging RocketChip}
To demonstrate the productivity gain of using \hgdb versus more traditional debugging techniques, we have reproduced
a known bug in the floating point unit (FPU) of the RocketChip, as shown in Listing~\ref{lst:bug-code-chisel}.
When the RocketChip simulator uses this code to execute a floating point comparison, the FPU output mismatches with the functional model. 

To debug the problem, we first use our IDE to set a tentative breakpoint on the floating point control logic where we need
to inspect the circuit state, as shown previously in Figure~\ref{fig:ide}. The breakpoint is set inside the \texttt{when}
statement, since this is the condition where floating-point comparison is enabled.

Once the breakpoint hits, we can examine the generator variables. The final output \texttt{toint} seems to be
correct but the exception flags are incorrectly set. We then examine the inputs to \texttt{dcmp}. \hgdb has the ability to reconstruct
structured variables from a list of flattened RTL signals; in this case, the IO ports are represented as
a Chisel \texttt{PortBundle}, as one would expect from the source code. With a quick glance, we can see that
\texttt{dcmp.io.signaling} is not set properly since it is permanently asserted.
It can be easily fixed by correcting \texttt{dcmp.io.signaling} assignment.

Although the bug seems obvious once we know where to look, it can be tricky and tedious if we start from the
generated RTL and waveform. Listing~\ref{lst:bug-code-rtl} shows (partially) the code regarding \texttt{toint}
calculation. It is difficult to discern the source for \texttt{toint} since the control flow is flattened by
the compiler.

\balance

\begin{lstlisting}[language=Scala,
  label=lst:bug-code-chisel,
  caption={Simplified Chisel source code containing an FPU bug. \texttt{dcmp} is a instance
  that compares two recoded floating-points.},
  captionpos=b,
  %numbers=left,
  %stepnumber=1,
  %firstnumber=853,
  breaklines=true,
  basicstyle=\footnotesize]
dcmp.io.a := in.in1
dcmp.io.b := in.in2
dcmp.io.signaling := Bool(true)
when (in.wflags) { // feq/flt/fle, fcvt
  toint := (~in.rm & Cat(dcmp.io.lt, dcmp.io.eq)).orR | (store >> minXLen << minXLen)
  io.out.bits.exc := dcmp.io.exceptionFlags
  intType := 0
}
\end{lstlisting}

\begin{lstlisting}[language=Verilog,
  label=lst:bug-code-rtl,
  caption={Generated RTL code that computes \texttt{toint}.},
  captionpos=b,
  basicstyle=\footnotesize]
wire [31:0] _toint_T = store[63:32];
wire [63:0] _toint_T_1 = {store[63:32], 32'h0};
wire [63:0] _GEN_36 = {{54'd0}, classify_out};
wire [63:0] _toint_T_2 = _GEN_36 | _toint_T_8;
wire [63:0] _GEN_23 = in_rm[0] ? _toint_T_2 : store;
wire [63:0] toint = in_wflags ? _GEN_29 : _GEN_23;
\end{lstlisting}


\subsection{Performance Benchmark}\label{sec:benchmark}
To demonstrate our system's minimal performance overhead, we have benchmarked the simulator performance using the
benchmark suite shipped with the RocketChip, running on a popular commercial simulator. The benchmark suite runs a set
of RISC-V programs and computes their clocks-per-instruction (CPI). We compare simulation speed on various
settings: 
\begin{itemize}
  \item \textbf{baseline} (optimized);
  \item \textbf{baseline + hgdb} (optimized);
  \item \textbf{debug} (unoptimized build and debug, without hgdb); and 
  \item \textbf{debug + hgdb} (unoptimized build and debug with hdgb).
\end{itemize}
The \hgdb runtime is compiled using \texttt{gcc}
with \texttt{-O3} optimization and we run the benchmark multiple times on an Intel Xeon 4214 CPU.

\begin{figure}[!htb]
  \centering
  \includegraphics[width=1\linewidth]{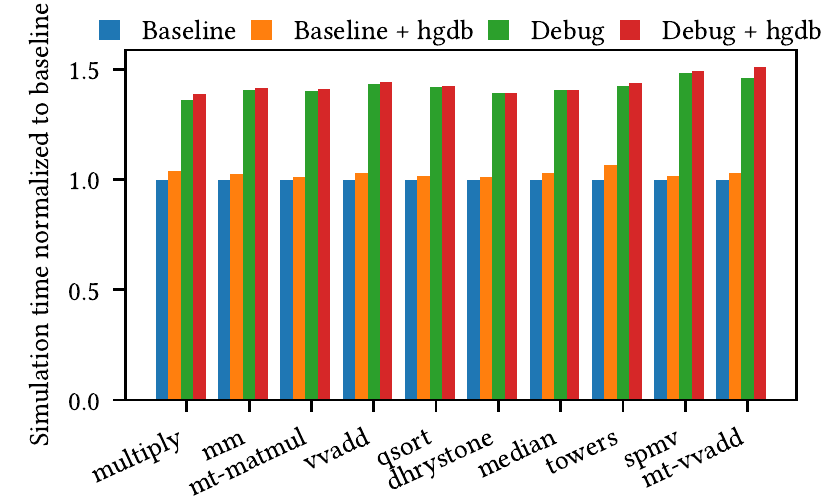}
  \caption{Benchmark performance for RocketChip under various testing conditions. Whether it is in baseline (optimized) or debug (unoptimized) mode, at no point does \hgdb overhead exceed 5\% of runtime.
  }\label{fig:benchmark-result}
  \Description[\hgdb benchmark result]{}
\end{figure}

As shown in Figure~\ref{fig:benchmark-result}, \hgdb introduces less than 5\% simulation overhead regardless of whether the design
is optimized or not. This is because the only overhead in \hgdb is the clock edge callback, which returns immediately. The
more complex the design, the more time the simulator spends to compute state updates. Hence the fixed cost of callback
per clock cycle is negligible. We observe similar minimal performance overhead on multiple commercial simulators and the open-source Verilator.

\section{Conclusions}
Debugging designs created by widely used HGFs is difficult, since  common source-level debugging techniques, such as symbol mapping and breakpoints, have previously not been available.
Our system addresses these issues, enabling hardware generated by
HGFs to be debugged at the source level. Since our system does not require the testing environment to be within the generator framework, it works with
any existing test bench and any SystemVerilog simulator. While this paper uses Chisel as a case study, the technique
can and has been applied to other HGFs. To encourage the conversation on HGFs source-level debugging, we
have open-sourced our debugging protocol as well as our implementation on Github.\footnote{\url{https://github.com/Kuree/hgdb}}

\bibliographystyle{./bibliography/ACM-Reference-Format}
\bibliography{bibliography.bib}


\begin{thebibliography}{12}


\ifx \showCODEN    \undefined \def \showCODEN     #1{\unskip}     \fi
\ifx \showDOI      \undefined \def \showDOI       #1{#1}\fi
\ifx \showISBNx    \undefined \def \showISBNx     #1{\unskip}     \fi
\ifx \showISBNxiii \undefined \def \showISBNxiii  #1{\unskip}     \fi
\ifx \showISSN     \undefined \def \showISSN      #1{\unskip}     \fi
\ifx \showLCCN     \undefined \def \showLCCN      #1{\unskip}     \fi
\ifx \shownote     \undefined \def \shownote      #1{#1}          \fi
\ifx \showarticletitle \undefined \def \showarticletitle #1{#1}   \fi
\ifx \showURL      \undefined \def \showURL       {\relax}        \fi
\providecommand\bibfield[2]{#2}
\providecommand\bibinfo[2]{#2}
\providecommand\natexlab[1]{#1}
\providecommand\showeprint[2][]{arXiv:#2}

\bibitem[\protect\citeauthoryear{Beamer and Donofrio}{Beamer and
  Donofrio}{2020}]%
        {ESSENT}
\bibfield{author}{\bibinfo{person}{Scott Beamer} {and} \bibinfo{person}{David
  Donofrio}.} \bibinfo{year}{2020}\natexlab{}.
\newblock \showarticletitle{Efficiently exploiting low activity factors to
  accelerate RTL simulation}. In \bibinfo{booktitle}{\emph{DAC}}.
  \bibinfo{publisher}{ACM/IEEE}, \bibinfo{address}{USA}, \bibinfo{pages}{1--6}.
\newblock


\bibitem[\protect\citeauthoryear{committee}{committee}{2017}]%
        {dwarf}
\bibfield{author}{\bibinfo{person}{The~DWARF committee}.}
  \bibinfo{year}{2017}\natexlab{}.
\newblock \bibinfo{title}{The {DWARF} Debugging Standard}.
\newblock
\newblock


\bibitem[\protect\citeauthoryear{Rosen, Wegman, and Zadeck}{Rosen
  et~al\mbox{.}}{1988}]%
        {ssa}
\bibfield{author}{\bibinfo{person}{Barry~K Rosen}, \bibinfo{person}{Mark~N
  Wegman}, {and} \bibinfo{person}{F~Kenneth Zadeck}.}
  \bibinfo{year}{1988}\natexlab{}.
\newblock \showarticletitle{Global value numbers and redundant computations}.
  In \bibinfo{booktitle}{\emph{Proceedings of the 15th SIGPLAN-SIGACT symposium
  on Principles of programming languages}}. \bibinfo{publisher}{ACM},
  \bibinfo{address}{USA}, \bibinfo{pages}{12--27}.
\newblock


\bibitem[\protect\citeauthoryear{Synopsys}{Synopsys}{2021}]%
        {verdi}
\bibfield{author}{\bibinfo{person}{Synopsys}.} \bibinfo{year}{2021}\natexlab{}.
\newblock \bibinfo{title}{Verdi}.
\newblock
\newblock


\bibitem[\protect\citeauthoryear{\textit{el.al}}{\textit{el.al}}{2018}]%
        {mamba}
\bibfield{author}{\bibinfo{person}{{S. Jiang} \textit{el.al}}.}
  \bibinfo{year}{2018}\natexlab{}.
\newblock \showarticletitle{Mamba: closing the performance gap in productive
  hardware development frameworks}. In \bibinfo{booktitle}{\emph{DAC}}.
  \bibinfo{publisher}{ACM/IEEE}, \bibinfo{address}{USA}, \bibinfo{pages}{1--6}.
\newblock


\bibitem[\protect\citeauthoryear{\textit{et al.}}{\textit{et al.}}{2017}]%
        {firrtl}
\bibfield{author}{\bibinfo{person}{A.~Izraelevitz \textit{et al.}}}
  \bibinfo{year}{2017}\natexlab{}.
\newblock \showarticletitle{Reusability is {FIRRTL} ground: Hardware
  construction languages, compiler frameworks, and transformations}. In
  \bibinfo{booktitle}{\emph{ICCAD}}. \bibinfo{publisher}{IEEE},
  \bibinfo{address}{USA}, \bibinfo{pages}{209–216}.
\newblock


\bibitem[\protect\citeauthoryear{\textit{et al.}}{\textit{et al.}}{1980}]%
        {high-level-language}
\bibfield{author}{\bibinfo{person}{David R~Ditzel \textit{et al.}}}
  \bibinfo{year}{1980}\natexlab{}.
\newblock \showarticletitle{Retrospective on high-level language computer
  architecture}. In \bibinfo{booktitle}{\emph{ISCA}}. \bibinfo{publisher}{ACM},
  \bibinfo{address}{USA}, \bibinfo{pages}{97–104}.
\newblock


\bibitem[\protect\citeauthoryear{\textit{et al.}}{\textit{et al.}}{2012}]%
        {chisel}
\bibfield{author}{\bibinfo{person}{J.~{Bachrach} \textit{et al.}}}
  \bibinfo{year}{2012}\natexlab{}.
\newblock \showarticletitle{Chisel: Constructing hardware in a Scala embedded
  language}. In \bibinfo{booktitle}{\emph{DAC}}. \bibinfo{publisher}{ACM/IEEE},
  \bibinfo{address}{USA}, \bibinfo{pages}{1212--1221}.
\newblock


\bibitem[\protect\citeauthoryear{\textit{et al.}}{\textit{et al.}}{2010}]%
        {genesis}
\bibfield{author}{\bibinfo{person}{Ofer~{Shacham} \textit{et al.}}}
  \bibinfo{year}{2010}\natexlab{}.
\newblock \showarticletitle{Rethinking digital design: Why design must change}.
\newblock \bibinfo{journal}{\emph{IEEE micro}} \bibinfo{volume}{30},
  \bibinfo{number}{6} (\bibinfo{year}{2010}), \bibinfo{pages}{9--24}.
\newblock


\bibitem[\protect\citeauthoryear{\textit{et al.}}{\textit{et al.}}{2011}]%
        {uvm}
\bibfield{author}{\bibinfo{person}{Young-Nam~Yun \textit{et al.}}}
  \bibinfo{year}{2011}\natexlab{}.
\newblock \showarticletitle{Beyond {UVM} for practical {SoC} verification}. In
  \bibinfo{booktitle}{\emph{International SoC Design Conference}}.
  \bibinfo{publisher}{IEEE}, \bibinfo{address}{USA}, \bibinfo{pages}{158--162}.
\newblock


\bibitem[\protect\citeauthoryear{\textit{et.al}}{\textit{et.al}}{2014}]%
        {calagar2014source}
\bibfield{author}{\bibinfo{person}{{N. Calagar} \textit{et.al}}.}
  \bibinfo{year}{2014}\natexlab{}.
\newblock \showarticletitle{Source-level debugging for FPGA high-level
  synthesis}. In \bibinfo{booktitle}{\emph{FPL}}. \bibinfo{publisher}{IEEE},
  \bibinfo{address}{USA}, \bibinfo{pages}{1--8}.
\newblock


\bibitem[\protect\citeauthoryear{Truong and Hanrahan}{Truong and
  Hanrahan}{2021}]%
        {magma}
\bibfield{author}{\bibinfo{person}{Leonard Truong} {and} \bibinfo{person}{Pat
  Hanrahan}.} \bibinfo{year}{2021}\natexlab{}.
\newblock \bibinfo{title}{Magma circuits}.
\newblock
\newblock
\urldef\tempurl%
\url{https://github.com/phanrahan/magma}
\showURL{%
\tempurl}


\end{thebibliography}

\end{document}